\documentclass[twocolumn]{aastex61}

\usepackage{graphicx}

\accepted{December 4, 2017}

\begin{document}

\title{Magnetospheric Gamma-Ray Emission in Active Galactic Nuclei}
\shorttitle{Katsoulakos \& Rieger}

\correspondingauthor{Grigorios Katsoulakos}
\email{gkats@mpi-hd.mpg.de, Frank.Rieger@mpi-hd.mpg.de}

\author{Grigorios Katsoulakos}
\affiliation{International Max Planck Research School for  Astronomy and Cosmic Physics, University of Heidelberg (IMPRS-HD)}
\affiliation{ZAH, Institut f\"ur Theoretische Astrophysik, Universit\"at Heidelberg, Philosophenweg 12,
69120 Heidelberg, Germany}
\affiliation{Max-Planck-Institut f\"ur Kernphysik, P.O. Box 103980, 69029 Heidelberg, Germany}
\nocollaboration

\author{Frank M. Rieger}
\affiliation{ZAH, Institut f\"ur Theoretische Astrophysik, Universit\"at Heidelberg, Philosophenweg 12,
69120 Heidelberg, Germany}
\affiliation{Max-Planck-Institut f\"ur Kernphysik, P.O. Box 103980, 69029 Heidelberg, Germany}
\nocollaboration

\begin{abstract}
The rapidly variable, very high-energy (VHE) gamma-ray emission from Active Galactic Nuclei (AGN) has been frequently associated with 
non-thermal processes occurring in the magnetospheres of their supermassive black holes. The present work aims to explore the adequacy 
of different gap-type (unscreened electric field) models to account for the observed characteristics. 
Based on a phenomenological description of the gap potential, we estimate the maximum extractable gap power $L_{gap}$ for different 
magnetospheric set-ups, and study its dependence on the accretion state of the source. $L_{gap}$ is found to be in general proportional 
to the Blandford-Znajek jet power $L_{BZ}$ and a sensitive function of gap size $h$, i.e. $L_{gap} \sim L_{BZ} (h/r_g)^{\beta}$, where 
the power index $\beta \geq 1$ is dependent on the respective gap-setup. The transparency of the black hole vicinity to VHE photons 
generally requires a radiatively inefficient accretion environment and thereby imposes constraints on possible accretion rates, and 
correspondingly on $L_{BZ}$. Similarly, rapid variability, if observed, may allow to constrain the gap size $h\sim c \Delta t$. Combining these 
constraints, we provide a general classification to assess the likelihood that the VHE gamma-ray emission observed from an AGN can be 
attributed to a magnetospheric origin. When applied to prominent candidate sources these considerations suggest that the variable (day-scale) 
VHE activity seen in the radio galaxy M87 could be compatible with a magnetospheric origin, while such an origin appears less likely for the 
(minute-scale) VHE activity in IC310.
\end{abstract}


\section{Introduction} \label{sec:01}
Non-thermal magnetospheric processes in the vicinity of supermassive black holes (BH) have been considered to provide a promising explanatory
framework for the origin of ultra-high energy cosmic rays (UHECRs) and the rapidly variable very high energy (VHE; $>100$ GeV) gamma-ray 
emission in AGN \citep[e.g., see][for review and references]{rie11}.

Accretion processes in AGN could support magnetic field strengths up to $B\sim 10^5$ Gauss on black hole horizon-scales $r_H$. A black hole 
embedded in such a magnetic field and rotating with angular frequency $\Omega^H =a_s c/2 r_H$, where $a_s$ is the dimensionless black hole
spin parameter, will induce an electric field of magnitude $E \sim (v_{rot}/c) \times B \sim  (\Omega^H r_H/c)~ B \sim (a_s/2)~B$ corresponding 
to a permanent voltage drop across the horizon of magnitude $\Delta\mathcal{V}_{gap} \sim E\,r_H \sim (a_s/2)r_H B$. This provides for the 
general possibility of particle acceleration and an efficient electromagnetic (Poynting flux) extraction of rotational energy from the black hole 
\citep[e.g.,][]{tho86,bes00}.\\
It has been proposed that the presence of strong electric fields close to the supermassive black holes in AGN could also facilitate the acceleration 
of cosmic rays to UHE energies of $\sim10^{20}$ eV, perhaps even after their main nuclear activity has ceased and the sources became quiescent
("dormant AGN") \citep[e.g.,][]{bol99,bol00,aha02,lev02,ner09,kal12,mon17}.
Proposals of this sort generally require that the electric field is not significantly shortened out (i.e., that a "vacuum gap" exists) and that curvature 
losses do not introduce a strong suppression \citep[cf.][]{ped11}. Similarly, direct electric acceleration of electrons could lead to curvature and inverse 
Compton emission in the VHE domain, and potentially trigger a pair cascade that could short out the electric field and lead to vacuum breakdown 
\citep{lev00}.

Magnetospheric gamma-ray emission has recently received a particular impetus in the context of the rapidly variable VHE emission detected from 
misaligned AGN, most prominently from the radio galaxies M87 ($d\simeq 16$ Mpc) and IC310 ($d\sim 80$ Mpc) \citep[e.g.,][]{ner07,vin10,lev11,
ale14,bro15,pti16,hir16,hir16b}. This goes along with the recognition that $r_g/c$, where $r_g=r_s/2=GM_{BH}/c^2$ is the gravitational radius of the 
black hole, provides a characteristic variability timescale of the emission in the frame of the galaxy, and that specific set-ups are needed to tap sufficient 
power on significantly shorter time scales \citep[e.g.,][]{bar10,gia10,aha17}. The day-scale VHE activity in M87 \citep[e.g.,][]{aha06,alb08,acc09} and 
the minute-scale one in IC310 \citep{ale14} correspond to scales of about $4\,r_g/c$ and $0.2\,r_g/c$, respectively. The VHE variability thus provides 
clear evidence for a compact emission zone, and this usually tends to be associated with smaller distance scales, i.e. a location closer to the black 
hole. Nevertheless, this does not necessarily have to be the case. In fact, the moderate angular resolution of VHE gamma-ray instruments ($\sim 
0.1^{\circ}$ does not allow to spatially resolve such scales, and this introduces some uncertainties. The situation can however be improved 
significantly by combining VHE with high-resolution radio observations capable of probing scales down to several tens of $r_g$. In the case of
M87 rapid VHE activity seems to precede the ejection of a radio-emitting component close to the black hole, suggesting that (at least sometimes)
the VHE emission originates in the vicinity of its black hole \citep{acc09}.

Misaligned AGN, with jets inclined such that Doppler factors are modest ($D \lesssim$ a few), belong to the most promising source class in this
regard, as possible Doppler modifications of observed timescales are reduced and the non-thermal jet radiation does not necessarily overpower 
the non-boosted magnetospheric emission \citep[cf.][]{rie17}. The observation of rapid variability on the other hand, imposes important constraints 
on possible gap sizes and extractable powers, and offers critical information as to the plausibility of relating the emission to a magnetospheric 
origin in a particular source. This is obviously of relevance for assessing the potential of different candidate sources. As we will show later on 
(Sec.~\ref{sec02}), however, different realisations of the gap potential are in principle conceivable. While this complicates the picture somewhat, 
observations can allow to favour/disfavour some of them, thereby offering additional clues on e.g. a possible disk-BH-jet connection.
Pair cascade processes in the black hole magnetosphere could possibly provide the plasma source required to ensure a force-free MHD outflow 
in a Blandford-Znajek type set-up \citep{bla77,lev11}. This in principle allows to link magnetospheric processes to accretion and jet formation, and 
makes magnetospheric studies of general relevance for jet investigations, and vice versa.

This paper is structured as follows. In Sec.~\ref{sec02} basic features of the black hole environment are presented, the gap potential is evaluated 
for different assumptions and the respective acceleration and radiative processes are analysed. This is then used to derive a constraint on the 
maximum possible gap power as a function of gap height. Application to the well-known VHE emitters M87 and IC310 is discussed in Sec.~\ref{sec:03}. 
The conclusions are shortly summarized in Sec.~\ref{sec:04}.

\section{Magnetospheric emission} \label{sec02}
Magnetospheric emission models usually rely on efficient "gap-type'' particle acceleration \citep[but see, e.g.,][for exemptions]{rie08,osm17}.
According to Ohm's law, $\textbf{J}=\sigma(\textbf{E}+\textbf{V}/c\times\textbf{B})$, the deficiency of electric charges (i.e., low conductivity 
$\sigma$) within the black hole magnetosphere can directly lead to the formation of (non-degenerate) regions with an unscreened parallel 
electric field component, i.e. $\textbf{E}\cdot\textbf{B}\neq 0$. Thus, magnetospheric particles moving along the magnetic fields into such 
charge-empty ("gap")  regions can be strongly accelerated to high energies by these parallel electric field components. 
Gaps can in principle occur under several conditions. Extended gaps (with sizes $h\geq r_{g}$), for example, are known for the vacuum black 
hole magnetosphere \citep{wal74}. Thinner gaps (with sizes $h<r_{g}$) can however exist as well and might also be expected in the context of 
degenerate, force-free outflows (i.e., in ideal MHD). As an example, we mention the null surface (located close to $r_g$) formed due to the 
frame-dragging effect \citep{bes92,hir98} and the stagnation surface (typically located at a few $r_g$) which divides MHD inflow from outflow 
regions \citep[e.g.,][]{glo14,bro15}. In these places, continuous charge replenishment (by particle creation or diffusion) has to occur in order to 
sustain the required Goldreich-Julian charge density $\rho_{GJ}=-e\,n_{GJ} \simeq -\Omega^F B_{\perp}/(2\pi c)$.\\  
Throughout this paper, we do not consider a specific gap position, but adopt a more phenomenological description. We assume that primary
particles can be injected into the magnetosphere by processes in the accretion environment (e.g., via annihilation of MeV photons emitted 
by a hot accretion flow). As has been shown elsewhere, the injected seed particle density is not sufficient for complete screening ($n_{\pm}
<n_{GJ}$) below a certain accretion rate $\dot{M}$ \citep{lev11}. This implies that gaps can appear if the accretion flow is advection-dominated 
(ADAF).

\subsection{The black hole vicinity} \label{ssec:2.1}
We consider a rotating black hole of spin parameter $a_s \simeq 1$ and mass $M_{BH}=M_{9}\times 10^9 M_{\odot}$ onto which gas accretion 
occurs. The central black hole is fed by the accretion flow at a rate $\dot{M}=\dot{m}\dot{M}_{Edd}$ expressed in Eddington units, where 
$\dot{M}_{Edd}=L_{Edd}/\eta_{c}c^2 \approx 1.4\times 10^{27} M_{9}$ g s$^{-1}$, $L_{Edd}=1.3 \times 10^{47}M_9$ erg $s^{-1}$ is the fiducial 
Eddington luminosity and $\eta_{c}\sim 0.1$ the canonical conversion efficiency. This corresponds to a reference limit on the accretion luminosity 
$L_{disk} = \eta_{c}\dot{M}c^2=\dot{m}L_{Edd}$ that is comparable to the one for a steady, standard (geometrically thin, optically thick) disk. 
As shown later on, however, a radiately inefficient accretion flow is a prerequisite for the escape and thus observability of magnetospheric VHE 
gamma-rays. For an optically-thin advection-dominated accretion flow (ADAF) in which most of the viscously dissipated energy is advected with 
the flow, cooling is almost inefficient, resulting in a much reduced luminosity $L_{ADAF}/L_{disk} \propto \dot{m} \ll 1$. ADAFs can only exists 
below a critical accretion rate, usually requiring $\dot{m}_c \simeq 0.4 \alpha_v^2 \lesssim 0.015$ (with $\alpha_v\lesssim 0.2$) \citep{mah97,nar98}, 
though more restrictive conditions ($\dot{m}_c\sim 0.003$) have been reported as well \citep[e.g.][]{bec02,yua14}.\\ 
We can approximate the characteristic disk magnetic field strength by assuming that the equipartition magnetic pressure is half the gas pressure, 
viz., $B^2/8 \pi= 0.5 \rho_i c_s^2$ where $c_s \simeq c\,(r_s/r)^{1/2}/\sqrt{3}$ is the sound speed. Though current simulations indicate some 
deviation from equipartition \citep[$\beta_{ADAF}>0.5$, cf.][]{yua14}, this will provide a useful upper limit. With $4\pi r^2\rho_i v_r=\dot{M}$ and 
radial inflow speed $v_r \simeq 0.5 \alpha_v v_f = (1/\sqrt{8}) \alpha_v (r_s/r)^{1/2} c$, where $\alpha_v$ is the viscosity coefficient, the disk 
magnetic field becomes $B_d \simeq 2.1 \times 10^4 \alpha_v^{-1/2} M_9^{-1/2}\dot{m}^{1/2} (r_s/r)^{5/4}$ G, which agrees well with previously 
reported ADAF results \citep[e.g.,][]{mah97,nar98,yi99}. Evaluating at characteristic radius $r=1.5 r_g$ \citep{mei01} and using $\alpha_v=0.1$
as reference value \citep[e.g.,][]{kin07,yua14}, the inner disk field could thus reach strengths of 
\begin{equation}\label{Badaf}
	B_d\approx 10^5 \dot{m}^{1/2}M_{9}^{-1/2} \quad \mathrm{G}.  \label{eq01}
\end{equation} Given that for ADAFs the disk scale height is about $H \sim r$, the strength of the poloidal magnetic field threading the black hole 
$B \simeq B_d \times (H/r)^n$, $n\sim 1$ \citep[][]{liv99,mei01}, is expected to be comparable in magnitude. In fact, taking the field-enhancing 
shear in the Kerr metric into account, the field threading the black hole may be a factor of about $2.3$ larger \citep{mei01}, i.e. $B_{d,h} \simeq 
2.3 B_d$, which would bring it close to the value inferred from GRMHD jet simulations in the context of magnetically arrested disks \citep[e.g.,]
[]{tch11,tch12,yua14}.

The emission spectrum of an ADAF is produced by synchrotron, inverse Compton and bremsstrahlung radiation of relativistic thermal electrons, 
and typically extends from radio frequencies up to hard X-rays \citep{nar95}. Like any disk emission, this radiation constitutes a potential target 
for any magnetospheric VHE $\gamma$-rays. For sufficiently small accretion rate $\dot{m}$ the peak energy $\epsilon_{s}$ and the luminosity 
$L_{s}$ of the soft photon field become \citep[viz. the synchrotron peak]{mah97}
\begin{equation}\label{ADAF_peak}
	 \epsilon_{s}\approx 0.2\, \dot{m}^{3/4}M_{9}^{-1/2}T_{e,9}^2 \quad \mathrm{eV}, \label{eq02}
\end{equation}
\begin{equation}
	 	L_{s}\approx 5\times 10^{43} \,\dot{m}^{9/4}M_{9}^{1/2}T_{e,9}^7 \quad \mathrm{erg\,s}^{-1}, \label{eq03} 
\end{equation}
where $T_{e,9}=T_{e}/10^{9}\approx 5$ is the characteristic electron temperature which depends weakly on the accretion rate $\dot{m}$ \citep{mah97} 
and the radial distance $r$ at the inner edge of the ADAF \citep{nar95,man97}. For simplicity, we thus fix the temperature to $T_{e,9}=5$ in all our 
following calculations. Correspondingly, the soft photon number density can be expressed as $n_{s}\simeq L_{s}/(2\pi r_{s}^2 c \epsilon_{s})\approx 
4 \times 10^{19}\dot{m}^{3/2}M_{9}^{-1}\quad \mathrm{cm}^{-3}$.

Accreting black hole systems are capable of ejecting powerful jets. On average the maximum power of these jets, $L_{jet}$, should be comparable 
to the available accretion power $\dot{M} c^2$ \citep[e.g.,][]{ghi14}. This should also hold for ergospheric-driven (BZ) jets as the magnetic flux carried 
onto the black hole is proportional to $\dot{M}$ (cf. eq.~[\ref{Badaf}]). GRMHD simulations in fact show that the jet power does not exceeds $\dot{M}c^2$ 
by more than a factor of $\sim3$ \citep[e.g.,][]{tch11}. Hence, on phenomenological grounds one could write $L_{jet} = \xi \dot{M} c^2  = \xi L_{disk}/\eta_c 
= (\xi/\eta_c)\, \dot{m} L_{Edd}$ with $\xi \lesssim 3$.
This could be related to the electromagnetic extraction of rotational energy of the supermassive black hole \citep{tho86}. In the case of a rotating, 
force-free black hole magnetosphere (i.e., efficient energy extraction) the maximum Blandford-Znajek (BZ) jet power is given by
\begin{eqnarray}
	L_{BZ}&=&\Omega^{F}(\Omega^{H}-\Omega^{F})B_{\perp}^2\frac{r_{H}^4}{c}
	             = (\Omega^F)^2 B_{\perp}^2 \frac{r_H^4}{c} \\ \nonumber
	             &=& \frac{1}{16} a_s^2 cr_H^2 B_{\perp}^2 \approx 2\times 10^{48}\,
	             \dot{m}\, M_{9} ~ \left(\frac{B_{\perp}}{B_{d,h}}\right)^2 \mathrm{erg\,s}^{-1},  \label{eq04}
\end{eqnarray}
where $\Omega^{H}= a_s\,c/2 r_{H}$ is the black hole angular velocity, $\Omega^{F}=\Omega^{H}/2$ is the angular velocity of the field lines in the 
case of efficient extraction, and  $B_{\perp} \simeq B_{d,h} \simeq 2.3 B_d$ is the strength of the normal magnetic field component threading the 
event horizon $r_H = r_g\,(1+\sqrt{1-a_s^2}) \sim r_g$ (for $a_s\sim1$). This concurs with the considerations above, $L_{BZ} = \xi \dot{M} c^2 
\lesssim 4\times 10^{48} \dot{m} M_9$ erg/s, and suggesting that the BZ jet power provides a useful measure for $\dot{M}$ and vice versa.
 
In general, magnetospheric VHE emission is perceived as a sub-product of an universal operation. If complete screening into the magnetosphere is 
not achieved ($\textbf{E}\cdot\textbf{B}\neq 0$), then particles, accelerated within the gaps, are likely to emit multiple VHE radiation \citep[e.g.,][]{lev00,
rie11}. Beyond an energy threshold, the VHE photons are absorbed by the ambient soft photon field producing secondary pairs. These will again be 
accelerated, their radiation being accompanied by further absorption/pair creation. In such a way, a photon-electron cascade is triggered, that 
guarantees a charge multiplicity such that force-freeness and MHD jet launching is ensured. Below the energy threshold, VHE photons can escape 
the black hole environment. Variable VHE emission observed from under-luminous AGN could thus signal the onset of relativistic jet formation
\citep{lev11}.       
  
\subsection{Gap electric field and potential} \label{ssec:2.2}
A quasi-steady magnetospheric gap can be formed in an under-dense environment ($n<n_{GJ}$). We consider in the following the gaps to be 
quasi-spherical, to be located at radial distance $R_{gap}$ close to $r_{g}$, and to possess a size or extension denoted by $h$. The voltage 
drop or gap potential then scales with the gap size depending on how the fields and boundaries are treated and different description are thus 
encountered. As gaps are often expected to be thin ($h\ll r_g$), this could lead to substantial differences.

\subsubsection{Heuristic constraint} \label{sssec:2.2.1}   
A heuristic constraint might be obtained from the global electric field of a force-free magnetosphere, i.e. $\textbf{E}=-\textbf{V}/c\times\textbf{B}$ 
in the high conductivity limit ($\sigma\rightarrow\infty$). Although the electric vector changes in the gap and only some part becomes parallel to 
the magnetic field lines, one may assume that its strength remains comparable in magnitude. Hence in order of magnitude, one could write for 
the electric field of the gap $E_{gap}\approx(V/c)B_{gap}\approx(\Omega^{F} R_{gap}/c) B_{gap}$ approximating $\sin\theta_b$ by unity. We
note that this expression in principle assumes that charge sheets or charge injection occurs just outside the gap boundaries (cf. eq.~[\ref{Gauss}] 
below). As this seems rather unexpected, the inferred $E_{gap}$ should be considered as providing a clear upper limit for possible realisations.

Noting $E_{gap}\sim\Delta\mathcal{V}_{gap}/h$, where $\Delta\mathcal{V}_{gap}$ denotes the voltage drop, and using $\Omega^{F}\approx 
\Omega^{H}/2$ and $\Omega^{H}=c/2r_{g}$, one would obtain
\begin{equation}\label{gap_constraint}
	\Delta\mathcal{V}_{gap}\approx \frac{1}{4}B_{gap} R_{gap}\left(\frac{h}{r_{g}}\right),  \label{eq05}
\end{equation} i.e. a voltage drop $\propto h$ scaling linearly with $h$ \citep[cf. in this context also][]{aha17}. Equation (\ref{eq05}) then represents 
a fiducial upper limit for the voltage drop of the magnetospheric gap that can be tapped for the acceleration of particles. For $B_{gap} \simeq 
B_{d,h} \simeq 2.3 B_d$ (eq.~[\ref{Badaf}]) and $R_{gap} \sim r_g$ this would give
\begin{equation}
	\Delta\mathcal{V}_{gap}\approx 2.5 \times 10^{21} \dot{m}^{1/2}M_{9}^{1/2}\left(\frac{h}{r_{g}}\right)\quad \mathrm{Volts},  \label{eq06}
\end{equation} noting that 1 Statvolt=300 Volts.

\subsubsection{Physical estimation} \label{sssec:2.2.2}
Different scaling laws of the voltage drop with gap size $h$ are in fact present in the literature. The apparent discrepancy can be related to 
different assumptions about the expected gap boundary conditions  \cite[see e.g.,][]{bes09}. In its simplest (one-dimensional, non-relativistic) 
form the gap electric field along $s$ in the presence of a non-zero charge charge density $\rho_e$ may be determined from Gauss' law
\begin{equation}\label{Gauss}
 \frac{dE_{||}}{ds}= 4\pi (\rho_e -\rho_{GJ})\,, 
\end{equation} and the electrostatic potential from 
\begin{equation}
\frac{d\Phi_e}{ds}= -E_{||}\,,
\end{equation} so that the voltage drop becomes $\Delta \mathcal{V}_{gap}= \Phi_e(s=h)-\Phi_e(s=0)$. 
   
To specify the electric field an appropriate boundary condition needs to be chosen. We may distinguish two cases, i.e. a highly (case {\it i}) and 
a weakly (case {\it ii}) under-dense one:\\ 
(i) Accordingly, case {\it i} can be characterized by $\rho_e \ll \rho_{GJ}$ and $E_{||}(s=0)\neq0$, with the developing pair cascade ensuring that 
the field gets screened at scale heigh $h$, i.e. $E_{||}(s=h)=0$. Thus, $dE_{||}/ds \approx -4\pi \rho_{GJ}$ and $E_{||} \approx -4\pi \rho_{GJ}\, s + 
const.$ (neglecting possible variations $\rho_{GJ}(s)$), so that using $E_{||}(s=h)=0$ one can write 
\begin{equation}
 E_{||}(s) = 4\pi \rho_{GJ} (h-s) = 4\pi\rho_{GJ} h \; \frac{ (h-s) }{h} = E_0 \frac{h-s}{h}\,,
 \end{equation}
with maximum $E_{||}(s=0)=E_0= 4\pi \rho_{GJ}\,h$. The voltage or potential drop then becomes
\begin{eqnarray}
\Delta \mathcal{V}_{gap} &=& \int_0^h E_{||}(s) ds =- 4\pi \rho_{GJ}\; \frac{h^2}{2}= -2\pi \rho_{GJ} \; h^2 \nonumber \\
                         &=& -2\pi \rho_{GJ}\, r_H^2 \; \left( \frac{h}{r_H} \right)^2
                         =  \Phi_0 \; \left( \frac{h}{r_H} \right)^2 \,,
 \end{eqnarray} 
where $\Phi_0 \equiv -2\pi \rho_{GJ} r_H^2 \simeq \Omega^F r_H^2 B_{\perp}/c = \Omega^H r_H^2 B_{\perp}/(2c)$, with $\rho_{GJ}\simeq -
\Omega^F B_{\perp}/(2\pi c)$, resulting in a scaling $\Delta \mathcal{V}_{gap} \propto h^2$ that is different to the one in eq.~(\ref{gap_constraint}). 
Such a scaling, $\propto h^2$, figures most promising in the gap context \citep[e.g.,][]{bla77,lev00,lev11}. We note that in this case some continuous
charge injection across the inner boundary would be usually needed to keep the gap quasi-steady lest it to become intermittent.  One could 
speculate that such a situation might arise close to the stagnation surface with the disk or corona facilitating plasma injection. Yet, given our 
limited understanding of the global magnetospheric structure for accreting black hole systems some caution needs to be exercised 
\citep[see e.g.,][]{con17}. It seems thus useful to address this problem also from another side, i.e., to employ observational constraints to 
evaluate the possible relevance of different gap realisations, and this approach is pursued below (Sec.~3).\\
(ii) A different dependence is, however, obtained for case {\it ii}. Here, $\rho_e \sim \rho_{GJ}$ and hence $E_{||}(s=0)\approx0$ with the cascade 
again ensuring that $E_{||}(s=h)=0$. Note that even if initially $\rho_e=\rho_{GJ}$ somewhere, deviations are expected as $\rho_{GJ} \propto 
\cos\theta_b$ varies along field lines. For non-trivial $E_{||}$ the chosen boundary condition can only be satisfied if $dE_{||}/ds$ changes sign
across the gap such that the electric field takes on an extremal value at $s=h/2$ (assuming symmetry), i.e., $dE_{||}/ds=0$ at $s=h/2$, which by 
Gauss' law implies $\rho(h/2)=\rho_{GJ}(h/2)$. We can thus Taylor-expand the charge-density term $(\rho_e- \rho_{GJ}) \equiv \rho_{eff}$ around 
$s=h/2$ to give $\rho_{eff}(s)= \rho_{eff}(h/2)+\frac{d\rho_{eff}}{ds}|_{(h/2)}\,(s-h/2)=\frac{d\rho_{eff}}{ds}|_{(h/2)}\,(s-h/2) \equiv \rho_{eff}'(h/2) \, 
(s-h/2)$ noting that $\rho_{eff}(h/2)=0$. Hence we have $dE_{||}/ds= 4\pi \rho_{eff}' \, (s-h/2)$ implying
\begin{equation}
 E_{||}(s) = 4\pi \rho_{eff}'\, \left( \frac{1}{2} s^2-\frac{h~s}{2}\right)
                                     = -2 \pi \rho_{eff}' r_H^2 \,\frac{s \,(h-s)}{r_H^2}\,.
\end{equation}
In the case of the null surface we approximately have \citep[e.g.,][]{hir16}
\begin{equation}\label{null_density}
\rho_{eff}'(h/2) \simeq - \frac{ \rho_{GJ} }{r_H} \simeq \frac{\Omega^F B_{\perp}}{(2\pi c r_H)}
\end{equation} 
and hence
\begin{eqnarray}\label{nu3_case}
\Delta \mathcal{V}_{gap}&=& \int_0^h E_{||}(s) ds =4\pi \rho_{eff}'\ \left[\frac{1}{6}s^3-\frac{1}{4} h s^2\right]_0^h \nonumber \\
                         &=& -4\pi \rho_{eff}' \frac{h^3}{12} 
                         = - \frac{1}{3} \pi \rho_{eff}' r_H^3 \; \left( \frac{h}{r_H} \right)^3\,,
\end{eqnarray} 
implying a scaling $\propto h^3$ that is different from those previously discussed. It seems tempting to associate such a case in the astrophysical
context with the region close to the null surface where the Goldreich-Julian charge density is expected to change sign. An analogous expression has 
been recently employed for studying the luminosity output of gaps in the AGN context \citep{hir16}. The strength of case (ii) lies in the fact that it 
seemingly provides for a transparent self-consistent realisation of a quasi-steady gap \cite[but cf. also][for caveats]{lev17}. 

To account for these differences and facilitate comparison, we employ a parametric expression in the following, i.e.
\begin{eqnarray}\label{eq07}
	\Delta\mathcal{V}_{gap}&=& \frac{1}{c}\, \eta \,\Omega^F  r_H^2 B_{\perp} \left(\frac{h}{r_H}\right)^{\nu} \nonumber \\ 
	&\approx&	 2.5 \times 10^{21} \eta ~\dot{m}^{1/2}M_{9}^{1/2}\left(\frac{h}{r_{g}}\right)^{\nu} \mathrm{Volts}\,, 
\end{eqnarray} where the respective sets of numerical parameters $\eta$ and $\nu$ are listed in Table~(\ref{tab:01}), and where $B_{gap}\simeq B_{d,h}$ 
has been assumed for the second expression on the right hand side. Given the employed approximation (cf. case ii), this expression formally applies 
to thin gaps, though we only demand $h \leq r_g$ in the following.
 
\begin{deluxetable*}{cccc}[ht]
\tablecaption{Gap voltage and luminosity parameters \label{tab:01}}
\tablecolumns{4}
\tablenum{1}
\tablewidth{0pt}
\tablehead{   \colhead{Exponent $\nu$}  &  \colhead{Coefficient $\eta$}  &  \colhead{Coefficient $\beta$}  &  \colhead{Reference}\\ }
\startdata
$1$ & $1$    &  1  & [1] \\
$2$ & $1$    &  2  & [2-4] \\
$3$ & $1/6$ &  4  & [5]  \\
\enddata
 \tablecomments{Parameters $\eta$ and $\nu$ for the voltage drop scaling $\Delta\mathcal{V}_{gap}\propto  \eta\,(h/r_H)^{\nu}$ as defined 
 in equation~(\ref{eq07}) and the corresponding maximum luminosity output $L_{gap} \propto L_{BZ}\, (h/r_H)^{\beta}$ following 
 equation~(\ref{gap_luminosity}). The first column specifies the power dependence on $h$, the second and third the corresponding values for 
 $\eta$ and $\beta$ respectively, and the fourth gives exemplary references, with  [1] = current work (reference limit), [2] \citet{bla77}, [3] \citet{lev00}, 
 [4] \citet{lev11}, [5] \citet{hir16}.}  
\end{deluxetable*}

\subsection{Particle acceleration and VHE radiation} \label{ssec:2.3}
A charged particle, injected into the magnetospheric gap, will be strongly accelerated. Consider an electron of energy $E_{e}=\gamma_{e}m_{e}c^2$ 
experiencing the potential drop of equation (\ref{eq07}). Its rate of energy gain per unit time is $d(\gamma_{e}m_{e}c^2)/dt=(c/r_{g}) e \Delta\mathcal{V}_{gap} 
\left(h/r_{g}\right)^{-1}$, where $\gamma_{e}$ is the Lorentz factor of the electron, $m_{e}$ is its rest mass and $c$ is the speed of light. 
This implies a characteristic acceleration time scale $\tau_{acc}=E_e/(dE_e/dt)$ of
\begin{equation}
	\tau_{acc}(\gamma_e)=10^{-12} \, \frac{\dot{m}^{-1/2}}{\eta} M_{9}^{1/2}\gamma_{e}\left(\frac{h}{r_{g}}\right)^{1-\nu} \quad \mathrm{s}\,.  
	\label{eq08}
\end{equation}
Electrons moving along field lines will, however, also experience losses due to field line curvature ("curvature radiation") and inverse Compton scattering 
\citep[e.g.,][for a review]{rie11}. Assuming that the curvature radius is roughly equal to the gravitational one (i.e., $R_{cur}\approx r_{g}$), the cooling time 
scale due to curvature radiation becomes
\begin{equation}
	\tau_{cur}(\gamma_e) \approx 4 \times 10^{30} \, M_{9}^2 \,\gamma_{e}^{-3} \quad \mathrm{s}.   \label{eq10}
\end{equation}
The whole magnetosphere as well as the gap are embedded in the ambient soft photon field of the disk (see subsection \ref{ssec:2.1}). Electrons 
undergoing acceleration along the field will thus also Compton up-scatter these soft photons to multiple VHE energies where the occurrence of 
$\gamma\gamma$-absorption can lead to the formation of a pair cascade \citep[e.g.,][]{lev11}. To explore the characteristic inverse Compton (IC) 
cooling time scale we approximate the soft photon field as isotropic and quasi-monoenergetic with $\epsilon_s$ and $L_s$ given by eqs.~(\ref{eq02}) 
and (\ref{eq03}). The IC electron cooling time scale then follows \citep{aha81}
\begin{eqnarray}
	\tau_{ic}(\gamma_e)=\frac{b^2}{3\sigma_{T} c n_{s}}\left[\left(6+\frac{b}{2}+\frac{6}{b}\right)\ln(1+b)-\ln^2(1+b)\right. \nonumber \\
	\nonumber \\
			\left.-2\, \mathrm{Li}\left(\frac{1}{1+b}\right)-\frac{\frac{11}{12}b^3+8b^2+13b+6}{(1+b)^2}\right]^{-1}\quad \mathrm{s}, \label{eq11}
\end{eqnarray}
where $\mathrm{Li}(x)\equiv-\int_{x}^{1}\ln y/(1-y) dy$ and  $b\equiv b(\gamma_e)=4\gamma_e\epsilon_s/(m_e c^2) \approx 3\times 10^{-5} 
\dot{m}^{3/4} \gamma_{e} M_{9}^{-1/2}$ is a non-dimensional quantity.

An electron will quickly reach its terminal Lorentz factor at which energy gain is balanced by radiative losses. Using appropriate values of  $\dot {m}$, 
$M_{9}$, $h$, $\nu$ and equating the acceleration time with the cooling ones (i.e., $\tau_{acc}=\tau_{cur}$ or/and $\tau_{acc}=\tau_{ic}$), the maximum
electron Lorentz factor $\gamma_{max}$ can be explored. It is worth emphasising that both radiation processes will take place, though the shortest 
one will impose the relevant constraint. Hence we may write $\gamma_{max}=\mathrm{min}(\gamma_{cur},\gamma_{ic})$, where the maximum 
particle Lorentz factor due to curvature is given by 
\begin{equation}
\gamma_{cur}\approx 4.5\times 10^{10} \eta^{1/4}\dot{m}^{1/8}M_{9}^{3/8}\left(\frac{h}{r_{g}}\right)^{\frac{\nu-1}{4}}\,.
\end{equation}

\begin{figure*}[t]
\epsscale{0.7}
\plotone{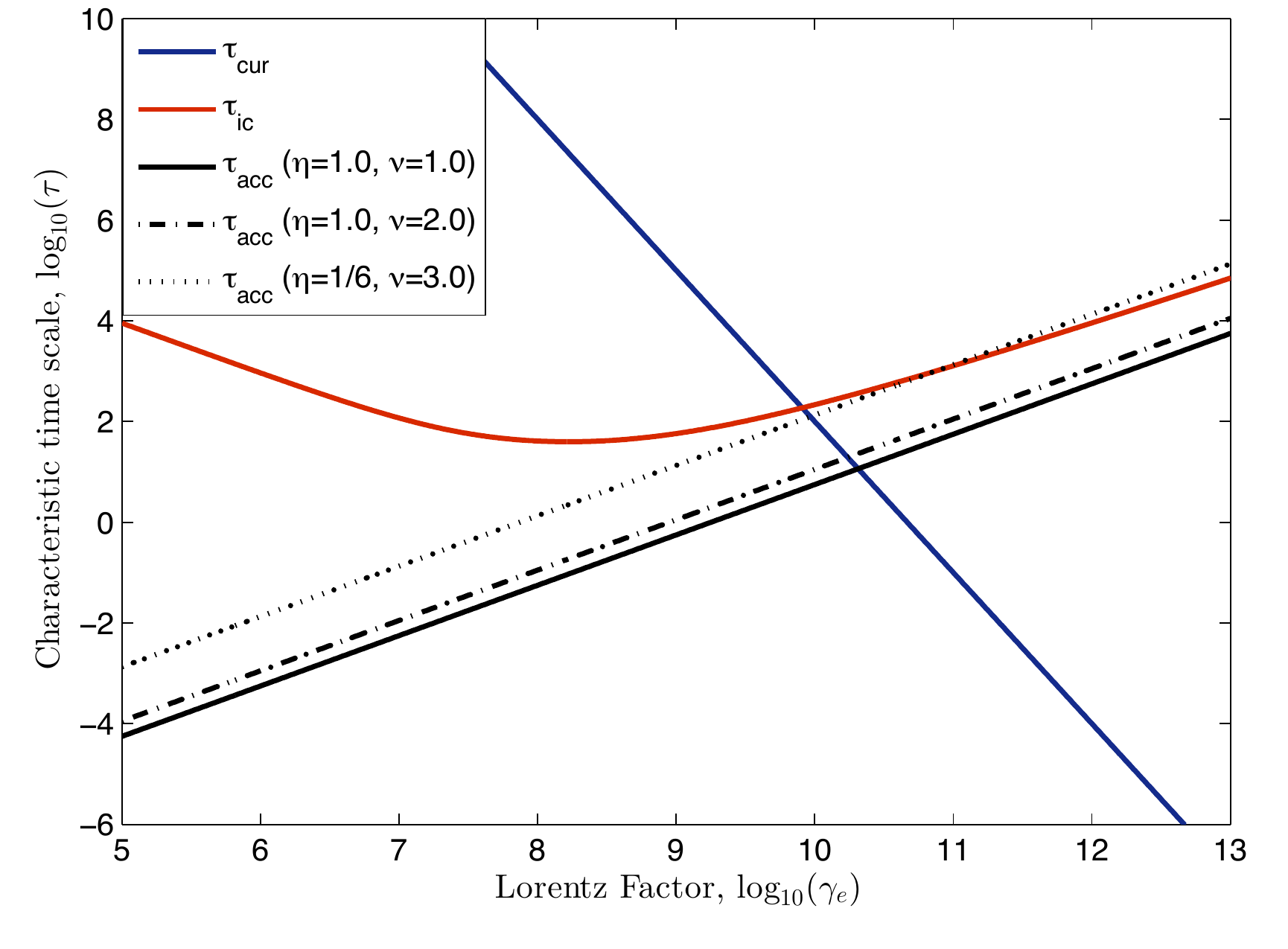}
\caption{Characteristic time scales as a function of Lorentz factor $\gamma_e$ for $M_{9}=5$, $\dot{m}=10^{-4}$, $h/r_{g}=0.5$. The solid blue 
line and red curve represent the time scales for  curvature and inverse Compton losses, respectively. The (rising) black lines represent the acceleration
time scales for the different gap potentials. The intersection points provide a measure of the achievable maximum energies.}\label{fig.1}
\end{figure*}

In figure~(\ref{fig.1}) the relevant time scales as a function of electron Lorentz factor are shown for a characteristic set of AGN parameters using the 
estimates in equations (\ref{eq08}), (\ref{eq10}) and (\ref{eq11}). For the considered choice, IC losses are weakened by Klein-Nishina effects and 
maximum particle energies for all considered voltage drops are essentially constrained by curvature losses. Note however, that for smaller gap sizes 
$h/r_g< 0.5$, IC losses will start to become of relevance and reduce achievable electron energies, foremost for $\nu =3$ (i.e. $\beta=4$). 
Figure ~(\ref{fig.1}) suggests that maximum Lorentz factors of $\gamma_{max}\sim 10^{10}$ could in principle be reached. This in turn implies that 
IC photons could reach multi-TeV energies, $\epsilon_{ic}=\gamma_{e}m_{e}c^2\leq 10^4$ TeV, while curvature emission could extend into the TeV 
regime, $\epsilon_{cur} = 3 c \hbar \,\gamma_{max}^3 / (2 R_{cur}) \sim  0.2\, (\gamma_{max}/10^{10})^3/M_9 $ TeV. Magnetospheric gaps in AGN 
are thus putative VHE emitting sites. If proton injection into the gap would occur (e.g., via diffusion), with Lorentz factors limited by curvature losses,
it is in principle conceivable that photo-meson (p$\gamma$) production in the soft photon field of the disk might contribute to this. 
    
\subsection{Gap luminosity} \label{ssec:2.4}
The radiative output of the gap depends on the number density of particles, $n_{\pm}$ undergoing acceleration and radiation in the gap. Gap closure
will occur once the effective charge density approaches the critical one, $|\rho_c|= e\,n_c $.  This allows for an estimate of the maximum achievable 
VHE gap power $L_{gap}$ given a voltage or potential drop $\Delta\mathcal{V}_{gap}$, 
\begin{equation}
L_{gap} \simeq n_c V_{gap} \frac{dE_e}{dt} \simeq -\frac{|\rho_c|}{e} (2\pi r_H^2 h)\frac{e~\Delta\mathcal{V}_{gap}~c}{h}\,,
\end{equation} approximating the relevant gap volume, $V_{gap}$, by a half-sphere of gap height $h$. The appropriate $\rho_c$ or $n_c$ to be employed
are dependent on the assumed gap set-up. For the heuristic and under-dense cases ($\nu=1$ and $\nu=2$ as delineated above, cf. eq.~[\ref{eq07}]), the 
critical value is typically comparable to the Goldreich-Julian density, i.e. $\rho_c =\rho_{GJ}\simeq -\Omega^F B_{\perp}/(2\pi c)=-\Omega^H B_{\perp}/(4\pi c)$.
In the weakly under-dense case ($\nu=3$), however, the appropriate value based on eq.~(\ref{null_density}) instead is $\rho_c \simeq \rho_{eff}' \,h \simeq 
\rho_{GJ}\,h/r_H$ \citep[e.g.,][]{hir16}. This results in scaling for the gap power $L_{gap} \propto h^4$ with power index increased by one compared to the 
respective potential drop (eq.~[\ref{nu3_case}]), while the index remains the same for the former cases.\\ 
Since $L_{gap} \propto (\Omega^F)^2 B_{\perp}^2$ we can also express the respective gap luminosity in terms of the Blandford-Znajek jet power, 
eq.~(\ref{eq04}). This gives
\begin{equation}\label{gap_luminosity}
L_{gap} = \eta \,L_{BZ} \left(\frac{h}{r_H}\right)^{\beta} \lesssim L_{BZ}\,,
\end{equation} where the respective sets of parameters are listed in Table~(\ref{tab:01}). For gap sizes $h \ll r_g$ then, the expected VHE output is much 
smaller than the jet power.

\subsection{Accretion environment} \label{ssec:2.5}
Independent of the preferred gap formation scenario, to avoid external $\gamma\gamma$-absorption of the gap-produced VHE gamma-rays 
and facilitate their escape from the vicinity of a supermassive black hole, magnetospheric models generally require an under-luminous or radiatively inefficient 
(RIAF) inner disk environment. For suppose that the disk would be of an (un-truncated) standard (geometrically thin, optically thick) type, for which the effective 
surface temperature obeys \citep[e.g.,][]{fra02}
\begin{eqnarray}
T_{\rm eff}(r) &=&\left[ \frac{3GM\dot{M}}{8\pi\sigma r^3}\left(1-\sqrt{\frac{r_{\rm in}}{r}}\right)\right]^{1/4} 
                     =  3.5 \times 10^5 \nonumber \\
                     &\times& \dot{m}^{1/4} M_9^{-1/4}  
                          \left(\frac{r_s}{r}\right)^{3/4}\left(1-\sqrt{\frac{r_{\rm in}}{r}}\right)^{1/4}
                                       \mathrm{K}\,,
\end{eqnarray} with $\sigma$ the Stefan Boltzmann constant. This profile peaks at a temperature $T_p \simeq 0.5~(3~GM \dot{M}/8\pi\sigma r^3)^{1/4}$ 
close to the inner disk radius $r\sim r_{in}\sim r_s$ and exhibits the canonical $r^{-3/4}$ dependence. For characteristic AGN parameters, the peak of the 
thermal disk emission, carrying a power of order $L_{disk} =\dot{m} L_{Edd}$, would then be occurring at eV energies, i.e., $\epsilon_p\simeq 2.8 kT_p 
\simeq 40~\dot{m}^{1/4} M_9^{-1/4}$ eV. This thermal disk radiation field would then provide an ideal target for the absorption of VHE photons producing 
pairs via $\gamma \,\gamma_{VHE} \rightarrow e^+e^-$. VHE photons of energy $\epsilon_{\gamma}$, in fact, interact most efficiently with ambient soft 
photons of energy 
\citep[e.g.,][]{rie11}
\begin{equation}\label{gamma_gamma}
     \epsilon_{d} \simeq 1~\left(\frac{1~\mathrm{TeV}}{\epsilon_{\gamma}}\right)~\mathrm{eV}\,.
\end{equation} The corresponding optical depth is of order $\tau_{\gamma\gamma} \sim \sigma_{\gamma\gamma }n_{d}~r$. With $\sigma_{\gamma\gamma} 
\simeq  0.2 \sigma_{T}$ and $n_{d} \sim L_{d}/4\pi r^2 c \epsilon_{d}$, approximating $L_{d} \simeq L_{disk}\times (\epsilon_{d}/\epsilon_p)^3$ (Rayleigh-Jeans limit) 
and $r\sim r_s$, this would be of the order of
 \begin{equation}
   	\tau_{\gamma\gamma} \sim 10^3 \, \dot{m}^{1/4} M_9^{3/4} \gg 1\,,
	\end{equation}
for VHE photons, i.e. greatly exceeding unity for conventional AGN parameters \citep[cf. also][]{zha97}. Hence, even if the black hole magnetosphere would 
produce VHE radiation via some gap mechanism, most of it is expected to become absorbed unless the disk would be of a radiatively inefficient (RIAF/ADAF) 
type where $L_{d}=L_{ADAF} \ll L_{disk}$ and the dominant part is emitted at energies much below 1~eV, cf. eq.~(\ref{ADAF_peak}). Very low accretion rates, 
or conservatively, the presence of a RIAF or ADAF thus becomes a necessary (yet not in itself sufficient) condition for the detectability of magnetospheric VHE 
emission in AGN. An ADAF-type (optically-thin) accretion flow with $H\sim r$ could also ensure the necessary poloidal magnetic field strength for efficient BZ 
power extraction \citep[e.g.,][]{liv99,mei01}.

\section{Application}\label{sec:03}
Gap-driven magnetospheric emission processes have been suggested as a potential generator of the highly variable VHE radiation seen from misaligned,
non-blazar AGNs \citep[e.g.,][]{ner07,lev11,ale14,vin15,pti16,hir16}. In this section, we seek to assess the potential of such scenarios on quite general
grounds, putting the relevant variability, transparency and power constraints in context.\\
First, VHE flux variability on short timescales $\Delta t$ will limit the size of the putative gap to $h \lesssim c \Delta t$. We generally expect $h$ to not 
exceed $ r_g$ if efficient pair cascade formation occurs. VHE flux variability thus imposes a limit on the extractable gap power as $L_{gap} \propto 
L_{BZ} (h/r_g)^{\beta}$, see eq.~(\ref{gap_luminosity}). On the other hand, to ensure transparency, i.e. for magnetospheric VHE emission to become 
observable, a radiatively inefficient disk (ADAF) environment is required (see subsection \ref{ssec:2.5}). This requires the accretion rate to satisfy $\dot{m} 
\lesssim \dot{m}_c$ and in turn leads to a constraint on the average jet power $L_{jet} \sim L_{BZ} \propto \dot{m}$, cf. eq.~(\ref{eq04}). 
Though a variety of values for $\dot{m}_c$ have been reported, $\dot{m}_{c} \sim 0.01$ appears to be a representative upper limit \citep[e.g.,][]{yua14}. 
Following this line of reasoning and assuming a rapidly spinning black hole, the constraints on the gap size and accretion rate thus translate into a 
characteristic upper-limit on the extractable VHE gap power of 
\begin{eqnarray}\label{max_power}
L_{gap}^{VHE} &\lesssim& 2\times 10^{46}~\eta~\left(\frac{\dot{m}_{c}}{0.01}\right)
                                                          ~\left(\frac{M_{BH}}{10^9 M_{\odot}}\right) \nonumber \\
                                                          &\times&
                                                          ~ \mathrm{min}\left\{\left(\frac{c\Delta t}{r_{g}}\right)^{\beta},1\right\} \; \mathrm{erg\,s}^{-1}\,.
\end{eqnarray} We note that this expression provides a quite general constraint and does not as yet presuppose a specific mechanism for pair injection. 
Under quasi-steady circumstances and provided that the flow is hot enough ($kT_e/m_ec^2 \sim1$), annihilation of MeV bremsstrahlung photons could 
well lead to a charge density in excess of $\rho_{GJ}$ before $\dot{m}$ approaches the critical value $\dot{m}_c$ \citep{lev11}. This would then introduce 
a further constraint on $\dot{m}$ as to the possible existence of a gap. For typical two-temperature models $T_e \gtrsim 100$ keV is expected \citep{yua14}.
However, uncertainties in the electron temperature $T_e$ caused by uncertainties in the electron heating parameters (in particular concerning the viscous 
dissipation that heats the electrons) along with flow intermittencies can introduce significant uncertainties in the pair production rate. At low accretion rates, 
early results, for example, suggested that $T_e\propto \dot{m}^{-q}$ with $0\lesssim q \lesssim 0.2$ \citep[e.g.,][]{mah97,esi97}, leading to some further
ambiguity concerning pair processes. Note that the scaling in equation~(\ref{max_power}) formally applies to small $h/r_g$ and that we have used $h\sim 
r_g$ for a representative upper limit. In principle a full general relativity model \citep[e.g.,][]{lev17} would be needed to self-consistently evaluate possible
gap widths.

In Fig.~(\ref{fig.2}), the product $P=10^{-48} L_{gap} M_{9}^{-1} (h/r_{g})^{-\beta}$ with $h\leq r_g$, which provides a normalized measure of the maximum 
gap power for a given black hole mass and gap size, is shown as function of accretion rate (or correspondingly, magnetic field strength threading the horizon, 
$B\propto \dot{m}^{1/2}$, cf. eq.~(\ref{eq01}). The case of a highly ($\eta=1$) and weakly ($\eta=1/6$) under-dense gap are given by the dotted and dash-dotted 
line, respectively. Observed VHE gamma-ray powers that are above these lines are unlikely to originate in (quasi steady) magnetospheric gaps. Both lines 
preserve their meaning for accretion rate lower than the critical one. These considerations are applied in the following to the most prominent candidate sources.

\subsection{M87} \label{ssec:3.1}
The Virgo Cluster radio galaxy M87 (NGC 4486), located at a distance of $d\simeq 16.7$ Mpc \citep{mei07} and believed to harbour a black hole of mass $M_{BH}
=(2-6)\times 10^9 M_{\odot}$, has been the first extragalactic source detected at VHE energies \citep{aha03}. Given its proximity, M87 has been a prime target to 
probe scenarios for the formation of relativistic jets with high-resolution radio observations exploring scales down to some tens of $r_g$ and much effort has been 
recently dedicated into this direction \citep[e.g.,][]{acc09,dol12,had14,had16,kin15,aki15,aki17}. At VHE energies, M87 has revealed at least three active $\gamma$-ray 
episodes during which day-scale flux variability (i.e., $h= c\Delta \tau_{obs} \sim r_{g}$) has been observed \citep{aha06,alb08,acc09,abr12,ali12}. The VHE 
spectrum is compatible with a relatively hard power-law (photon index $\sim 2.2$) extending from $300$ GeV to beyond $10$ TeV, while the corresponding TeV 
output is relatively moderate, with an isotropic equivalent luminosity of $L_{VHE}\simeq (3-10) \times 10^{40}$ erg\,s$^{-1}$. The inner, pc-scale jet in M87 is considered 
to be misaligned by $i \sim (15-25)^{\circ}$, resulting in modest Doppler boosting of its jet emission and creating challenges for conventional jet models to account 
for the observed VHE characteristics \citep[see e.g.,][for review and references]{rie12}. Gap-type emission models offer a promising alternative and different 
realisations have been proposed in the literature \citep[e.g.,][]{ner07,lev11,vin15,bro15,pti16}. M87 is overall highly under-luminous with characteristic estimates 
for its total nuclear (disk and jet) bolometric luminosity not exceeding $L_{\rm bol} \simeq 10^{42}$ erg\,s$^{-1}$ by much \citep[e.g.,][]{owe00,why04,pri16}, suggesting 
that accretion onto its black hole indeed occurs in a non-standard, advective-dominated (ADAF) mode characterized by an intrinsically low radiative efficiency 
\citep[e.g.,][]{dim03,nem14}, with inferred accretion rates possibly ranging up to $\dot{m} \sim10^{-4} \ll \dot{m}_c$ \citep[e.g.,][]{lev11} and black hole spin parameter 
close to its maximum one \citep[e.g.,][]{fen17}. For these values of the accretion rate, the soft photon field, cf. eqs.~(\ref{eq02}) and (\ref{eq03}), is sufficiently 
sparse, so that the maximum Lorentz factor $\gamma_{e}\sim 10^{10}$ of the magnetospheric particles is essentially determined by the curvature mechanism. 
The observed VHE variability is in principle compatible with $h\sim r_g$, so that the different dependence of the gap power on $\beta$, eq.~(\ref{gap_luminosity}), 
does not necessarily (in the absence of other, intrinsic gap closure considerations) imply a strong difference in the extractable gap powers. Fig.~(\ref{fig.2}) shows 
a representative point for M87 (taking $\beta=1$). The observed VHE luminosity of M87 is some orders of magnitudes lower than the maximum possible gap power 
(given by the dotted line) and within the bound imposed by ADAF considerations (vertical line). The observed VHE flaring events thus appear consistent with a 
magnetospheric origin. VLBI observations of (delayed) radio core flux enhancements indeed provide support for the proposal that the variable VHE emission in M87 
originates at the jet base very near to the black hole \citep[e.g.,][]{acc09,bei12,had12,had14}.

\subsection{IC 310}
The Perseus Cluster galaxy IC310 (J0316+4119) has revealed remarkable VHE variability during a strong flare in November 2012, exhibiting VHE flux variations 
on timescales as short as $\Delta t\simeq 5$ min \citep{ale14}. IC310 is located at a distance of $d\sim80$ Mpc (z=0.019) and widely believed to harbour a black 
hole of mass $M_{BH} \simeq 3 \times 10^8 M_{\odot}$ \citep[][but cf. also \cite{ber15} for a ten times smaller estimate]{ale14}. The flare spectrum in the energy 
range 70 GeV to 8.3 TeV appears compatible with a hard power law of photon index $\Gamma \simeq 2$ and does not show indications for internal absorption. 
The observed VHE fluxes can reach levels corresponding to an isotropic-equivalent luminosity of $L_{VHE} \simeq 2 \times 10^{44}$ erg\,s$^{-1}$. A variety of 
considerations based on the orientation of the jet in IC310 (probably $i\sim 10-20^{\circ}$), on kinetic jet power and timing constraints has led \citet{ale14} to 
disfavour alternative proposals for fast VHE variability, such as jet-star interaction \citep[e.g.,][]{bar12} or magnetic reconnection \citep[e.g.][]{gia13}. Detailed 
investigation, however, suggests that this does not have to be the case \citep[cf.][for details]{aha17}. Nevertheless, the fact that the VHE flux varies on timescales 
$\Delta t$ much shorter than the light travel time across black hole horizon scales, $r_g(3\times10^8 M_{\odot})/c = 25$ min, has been interpreted as evidence for 
the occurrence of gap-type particle acceleration on sub-horizon scales, i.e. of gap height $h\simeq 0.2 r_g$ \citep[e.g.,][]{ale14,hir16}. To sustain a steady, isotropic 
equivalent luminosity of $L_{bol} \sim 10^{44}$ erg\,s$^{-1}$, the average jet power should satisfy $L_j \gtrsim 10^{42}~(\theta_j/0.3~\mathrm{rad})^2$ erg\,s$^{-1}$ 
\citep[cf. also,][]{sij98,ahn17}, where $\theta_j$ denotes the jet opening angle, suggesting that typical accretion rates should exceed $\dot{m} \gtrsim 10^{-5} M_8^{-1}$ 
(where $M_8=M_{BH}/10^8 M_{\odot}$). Taking such a jet power as a reference, the expected gap emission would strongly under-predict the required VHE luminosity, 
see eq.~(\ref{gap_luminosity}). 
As $L_{gap}^{VHE} \propto B^2$ the gap would need to be temporarily threaded by much higher magnetic fields, and accordingly require much higher accretion rates,
cf. eq.~(\ref{Badaf}). If $h\sim r_g$ accretion rates of the order of $\dot{m}\sim 10^{-3}$ might seemingly be sufficient. The variability constraint $h\sim 0.2 r_g$,
however, implies that $L_{gap}^{VHE} \lesssim 6 \times 10^{45} \eta~(\dot{m}_c/0.01)~(0.2)^{\beta}$ erg\,s$^{-1}$, see eq.~(\ref{max_power}). This results in 
$L_{gap}^{VHE} \lesssim 2 \times 10^{44}$ erg\,s$^{-1}$ for the $\beta=2$ with $\eta=1$, and $L_{gap}^{VHE} \lesssim 10^{42}$ erg\,s$^{-1}$ assuming $\beta=4$ 
with $\eta=1/6$, see table~\ref{tab:01}.
In Fig.~(\ref{fig.2}), two representative points (i.e., for $\beta=1$ and $\beta=4$) illustrating the range for IC310 are shown. While the case with $\beta=2$ may 
appear marginally possible, we note that for accretion rates approaching $\dot{m}_c$ the existence of a gap is not guaranteed. In fact, annihilation of ADAF 
bremsstrahlung photons is likely to lead to an injection of pairs in excess of the Goldreich-Julian density, $n_{\pm}/n_{GJ} \sim 3\times 10^{12}\dot{m}^{7/2}$ 
\citep[][eq.~(8)]{lev11}, suggesting that rates $\dot{m} \lesssim 3 \times 10^{-4}$ are needed to avoid gap closure. Even for the most optimistic case ($\beta=1$), 
the maximum gap power, eq.~({\ref{max_power}), is then not expected to exceed $L_{gap}^{VHE} \sim 4 \times 10^{43}$ erg\,s$^{-1}$. We note that this concurs 
with a similar estimate in \citet{aha17}. This would imply that the noted VHE flaring event cannot be of a gap-type magnetospheric origin independent of the assumed 
power index $\beta$. A putative way out could be to assume an inner electron temperature $kT_e/(m_e c^2) <1$ such that bremsstrahlung emission would be 
suppressed at MeV energies thereby possibly relaxing the pair-injection constraint on $\dot{m}$. Whether this is feasible in the case of IC310 would need detailed disk 
modelling. But beside of this, the apparently huge magnetic fields required to thread the gap ($B_{d,h}\sim 2\times 10^5$ G in the case of $\beta=4$) would suggest 
a temporary increase in accretion rate in excess of $\dot{m}_{c}$ required for the existence of an ADAF. This, then, would make it again unlikely that 
magnetospheric VHE emission from IC310 should become observable, see Sec.~\ref{ssec:2.5}. The situation might, however, be more complex as unsteady
accretion and strong intermittency could be occurring. In the case of IC310 the $\sim5$ min flare event detected during 3.7 hrs of MAGIC observations in the night 
November 12-13, 2012, however, seems part of a higher source state, probably (considering earlier 2009/10 results) of day-scale duration $t_d$ \citep{ale14,ale14b,
ahn17}. If one takes the timescale $t_{th}$ for re-adjustment of thermal equilibrium as characteristic measure, $t_{th}\sim t_{dyn}/\alpha_v \sim 1/(\alpha_v~\Omega_k) 
\sim \alpha_v^{-1} (r/r_g)^{3/2}~r_g/c \sim 4~(r/r_g)^{3/2}$ hr $\lesssim t_d$, then this could be occurring sufficiently fast to ensure re-adjustment of the innermost 
disk parts. Another situation would be arising if the black hole mass in IC310 would indeed be smaller by an order of magnitude, i.e. $M_{BH} \sim3 \times 10^{7} 
M_{\odot}$, as suggested by \citet{ber15}. The observed rapid VHE variability of $\Delta t \sim 5$ min would then only imply $h\sim r_g$, such that the different 
dependence on $\beta$, e.g. eq.~(\ref{max_power}), would not make a significant difference. The limit introduced by eq.~(\ref{max_power}) would then suggest
that the observed VHE output might be formally achieved (assuming $\dot{m} \sim 0.01$). However, for such a rate and given the small black hole mass, 
eq.~(\ref{eq02}) would imply a synchrotron peak around $\epsilon_s \sim 1$ eV with associated power $L_s$ such, that the VHE photons are unlikely to escape 
absorption. In summary, though one could speculate that on $r_g$-scales the accretion flow evolves in a highly turbulent way, thereby changing its radiative 
characteristics, a gap-driven magnetospheric origin for the recent VHE flaring event in IC310 appears to be disfavoured unless its black hole mass and 
accretion state would be highly different.

\begin{figure*}[t]
\epsscale{0.7}
\plotone{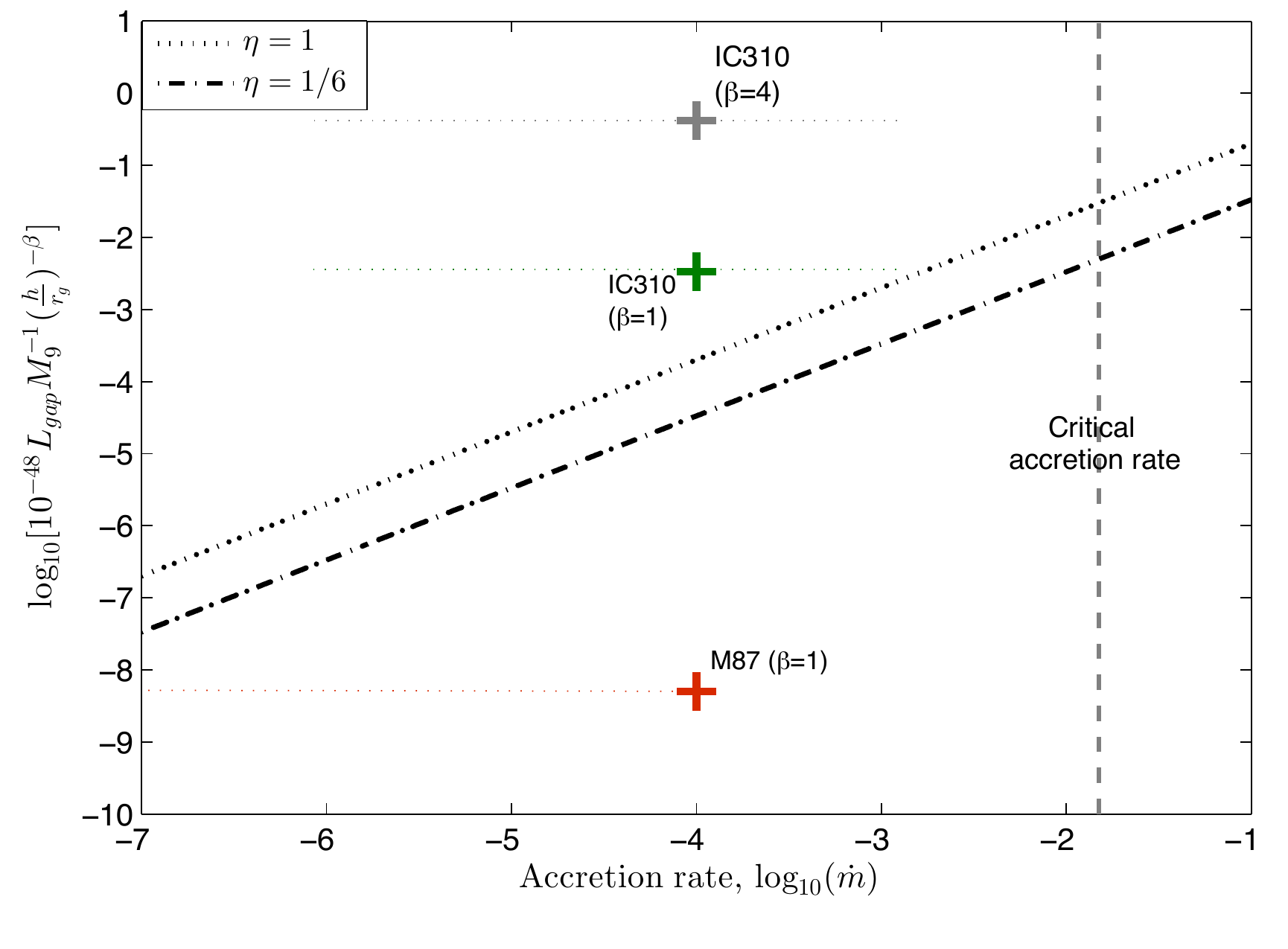}
\caption{Characteristic maximum power of a magnetospheric gap as a function of accretion rate $\dot{m}$. The dotted and dash-dotted lines 
represent the maximum for $\eta=1$ and $\eta=1/6$, respectively. Observed gamma-ray powers below these limiting lines could in
principle be produced by magnetospheric gaps.}\label{fig.2}
\end{figure*}

\section{Conclusion} \label{sec:04}
Gap-driven magnetospheric gamma-ray emission from rotating supermassive black holes is a potential candidate for the origin of the highly variable VHE 
emission seen in some AGN \citep[e.g.,][for review]{rie11}. The presence of strong, unscreened parallel electric field components on black hole horizon scales 
$\sim r_g$ could easily facilitate efficient particle acceleration, with the accompanying curvature and inverse Compton processes resulting in appreciable emission 
at gamma-ray energies. The efficiency and extractable power, however, depend on details of the gap set-up and different realizations are in principle conceivable 
and encountered in the literature. The present work explores possible implications of this by means of a simple phenomenological description that though heuristic, 
recovers the relevant dependencies. Accordingly, the maximum extractable gap power is in general proportional to the classical Blandford-Znajek jet power, $L_{BZ} 
\propto \dot{m}\,M_{BH}$, and a function of the gap height $h$, $L_{gap} \sim L_{BZ}\,(h/r_g)^{\beta}$, where the power index $\beta$ is dependent on the 
respective gap-setup (cf. Table~\ref{tab:01}). In order for this emission to become observable, VHE photons need to be able to escape the accretion environment 
of the black hole. Transparency to $\gamma\gamma$-pair production in fact requires an under-luminous or radiatively inefficient environment (RIAF/ADAF), and 
this introduces a relevant constraint on possible accretion rates of $\dot{m}\lesssim 0.01$. While for a fixed background a larger black hole mass (size) could be 
conducive to dilution of the soft photon field (facilitating VHE transparency) and increase $L_{BZ}$, the detection of rapid gamma-ray variability with $c \Delta t \sim 
h < r_g$ reduces the maximum gap power $L_{gap}^{VHE} \propto M_{BH}^{1-\beta}$ (for $\beta > 1$) and diminish the VHE prospects for source detection. When 
put in the context of current observations, these considerations suggest that the variable (day-scale) VHE activity seen in the radio galaxy M87 ($M_{BH} \simeq [2-6] 
\times 10^9 M_{\odot}$) may be compatible with a magnetospheric origin, while such an origin seems less likely for the (minute-scale) VHE activity in IC310 
(assuming $M_{BH} \simeq 3 \times 10^8 M_{\odot}$).\\
Our analysis implies that variability information will be crucial to get deeper insights into the physics of the putative gaps, to probe different potential scalings and 
to generally assess the plausibility of a magnetospheric origin. On average, however, (quasi steady) magnetospheric VHE gamma-ray emission in AGN is expected 
to be of rather moderate luminosity when compared to the strongly Doppler-boosted jet emission from blazars. The jet usually needs to be sufficiently misaligned for
the gap emission not to be masked by Doppler-boosted jet emission, making nearby misaligned AGN the most promising source targets. The possible impact of 
intermittencies in the gap formation process \citep[e.g.,][]{lev17} on the VHE characteristics could be of particular interest in this regard. The increase in sensitivity 
with the upcoming CTA instrument will allow to probe variability timescales  $< r_g/c$ for a number of these sources, and thereby allow a better census of
magnetospheric gamma-ray emitter.

\acknowledgments{Acknowledgments}

GK kindly acknowledges support by an IMPRS Fellowship. FMR kindly acknowledges financial support by a DFG Heisenberg Fellowship (RI 1187/4-1). 
We are grateful to A. Levinson for comments on early version of the manuscript. Discussions with F. Aharonian, K. Mannheim, C. Fendt, D. Khangulyan 
and D. Glawion are gratefully acknowledged.



\end{document}